\documentclass[12pt,preprint,preprint,nofootinbib]{revtex4}
\pdfoutput=1
\usepackage{epsf, array, color}
\usepackage{graphicx}
\usepackage{subfigure}
\usepackage{bbold}
\usepackage{amsmath}
\usepackage{amssymb}
\usepackage{mathtools}
\usepackage{epigraph}
\usepackage[utf8]{inputenc}
\usepackage{graphicx}
\usepackage[linktocpage,colorlinks,urlcolor=blue]{hyperref}
\usepackage{textcomp}
\usepackage{lscape}
\usepackage{pdflscape}

\usepackage{url}

\usepackage{dcolumn}% Align table columns on decimal point

\begin{document}

\def\fb{\rm fb}
\def\etal{{\it et al.}}
\newcommand{\bea}{\begin{eqnarray}}
\newcommand{\eea}{\end{eqnarray}}
\newcommand{\nn}{\nonumber \\}
\newcommand{\lag}{\ensuremath{{\cal L}}} 
\newcommand{\Tr}{\operatorname{Tr}}
\newcommand{\Log}{\operatorname{Log}}
\newcommand{\Det}{\operatorname{Det}}
\def\beq{\begin{equation}}
\def\eeq{\end{equation}}
\newcommand{\vev}[1]{\langle {#1} \rangle}
\newcommand{\lsim}{\lesssim}
\newcommand{\ord}[1]{\mathcal{O}{(#1)}}
\newcommand{\co}{\mathcal{O}}
\newcommand{\gsim}{\gtrsim}
\newcommand{\eq}[1]{Eq.~(\ref{#1})}

\newcommand{\nc}{\newcommand}

\newcommand{\C}{\mathcal{C}}
\newcommand{\Op}{{\cal O}}
\nc{\vp}{\phi}
\nc{\tvp}{\widetilde{\phi}}
\nc{\vpj }{\mbox{${\vp^\dag i\,\raisebox{2mm}{\boldmath ${}^\leftrightarrow$}\hspace{-4mm} D_\mu\,\vp}$}}
\nc{\vpjt}{\mbox{${\vp^\dag i\,\raisebox{2mm}{\boldmath ${}^\leftrightarrow$}\hspace{-4mm} D_\mu^{\,a}\,\vp}$}}

\def\nn{\nonumber}
\def\gev{\rm GeV}
\def\tev{\rm TeV}
\def\mev{\rm MeV}
\def\ev{\rm eV}
\def\met{\rm MET}
\def\br{{\tt Br}}
\def\higgs{{\rm Higgs}}
\newcommand{\vLam}{{v^2\over\Lambda^2}}
\newcommand{\TphiWB}{{\C_{\phi WB}}}
\newcommand{\Tll}{\C_{ll}}
\newcommand{\Tlu}{\C_{lu}}
\newcommand{\Tld}{\C_{ld}}
\newcommand{\Tle}{\C_{le}}
\newcommand{\Ted}{\C_{ed}}
\newcommand{\Tqe}{\C_{qe}}
\newcommand{\Teu}{\C_{eu}}
\newcommand{\Tee}{\C_{ee}}
\newcommand{\Tdd}{\C_{dd}}
\newcommand{\Tuu}{\C_{uu}}
\newcommand{\Tfe}{\C_{\phi e}}
\newcommand{\Tfd}{\C_{\phi d}}
\newcommand{\Tfu}{\C_{\phi u}}
\newcommand{\TuB}{\C_{uB}}
\newcommand{\TuW}{\C_{uW}}
\newcommand{\TW}{\C_{W}}
\newcommand{\Tfla}{\C_{\phi l}^{(1)}}
\newcommand{\Tfqa}{\C_{\phi q}^{(1)}}
\newcommand{\Tfqc}{\C_{\phi q}^{(3)}}
\newcommand{\Tlqa}{\C_{lq}^{(1)}}
\newcommand{\Tlqc}{\C_{lq}^{(3)}}
\newcommand{\Tqqa}{\C_{qq}^{(1)}}
\newcommand{\Tqqc}{\C_{qq}^{(3)}}
\newcommand{\TphiD}{\C_{\phi D}}
\newcommand{\Tflc}{\C_{\phi l}^{(3)}}
\newcommand{\TphiB}{\C_{\phi B}}
\newcommand{\TphiW}{\C_{\phi W}}
\newcommand{\Tphik}{\C_{\phi \square}}
\newcommand{\Tuda}{\C_{ud}^{(1)}}
\newcommand{\Tqda}{\C_{qd}^{(1)}}
\newcommand{\Tqua}{\C_{qu}^{(1)}}

%-----------------------------------
% Preprint Number
%-----------------------------------
\

\title{Flavorful Electroweak Precision Observables in the Standard Model Effective Field Theory}

\author{Sally Dawson$^{\, a}$ and Pier Paolo Giardino$^{\, b}$ }
%\email[]{}
%\author{}
%\author{}
\affiliation{
\vspace*{.5cm}
  \mbox{$^a$Department of Physics,\\
  Brookhaven National Laboratory, Upton, N.Y., 11973,  U.S.A.}\\
 \mbox{$^b$Instituto Galego de F\'isica de Altas Enerx\'ias, Universidade de Santiago de Compostela,}\\ \mbox{15782 Santiago de Compostela, Galicia, Spain}
 \vspace*{1cm}}

\date{\today}

\begin{abstract}
Electroweak precision observables (EWPO) measured at the $W$ and $Z$ poles provide stringent limits on possible beyond the Standard Model physics scenarios.  In an effective field theory  (EFT) framework, the next-to-leading order QCD and electroweak results for EWPO yield indirect limits on possible 4-fermion operators that do not contribute to the observables at tree level.  Here we calculate the next-to-leading corrections to EWPO induced by flavor non-universal 
4-fermion interactions  and find that the extracted limits on EFT coefficients have a strong dependence on the flavor structure of the 4-fermion operators. 
\end{abstract}

\maketitle

\section{Introduction}

Historically, measurements of electroweak precision observables  (EWPO) at the $W$ and $Z$ poles
have provided strong bounds on possible beyond the Standard Model (BSM) physics assumed to occur at some high scale $\Lambda$.  Similarly, in the Standard Model effective field theory (SMEFT) approach, comparisons
of EWPO  with theoretical predictions severely constrain the allowed values of the coefficients of the
effective field theory operators\cite{Dawson:2019clf,Corbett:2017qgl,Berthier:2015gja,Corbett:2021eux,deBlas:2019rxi,Dawson:2020oco}.  These bounds rely on precision calculations of the theoretical
expectations both in the Standard Model (SM) and in the SMEFT.
SMEFT calculations at next-to-leading order (NLO) QCD can be automated\cite{Degrande:2020evl}, while a growing number of SMEFT
electroweak NLO results exist\cite{Cullen:2020zof,Cullen:2019nnr,Gauld:2016kuu,Hartmann:2015aia,Hartmann:2015oia,Dawson:2018liq,Dedes:2018seb,Dawson:2018pyl,Dedes:2019bew,Dawson:2019clf,Hartmann:2016pil,Boughezal:2019xpp,Dawson:2018dxp,Dawson:2021xea}.
In a previous work\cite{Dawson:2019clf}, we computed the ${\cal{O}}({M_Z^2\over \Lambda^2})$  NLO QCD and electroweak corrections to the SMEFT predictions for EWPO at the $Z$ and $W$ poles.  At NLO, a dependence on operators beyond those occurring at tree level  in the analysis of EWPO  arises and limits can be placed on the corresponding coefficients.  
Quark and lepton flavor effects were neglected  in our earlier study and a $U(3)^5$ global flavor symmetry assumed for
all operators involving fermions.  Experimental anomalies in $B$ physics, however,  suggest that including flavor dependencies  in the  SMEFT analyses could be relevant\cite{Bruggisser:2021duo,Alasfar:2020mne}.  
 Furthermore, tree level global analyses of weak scale processes  have a significant dependence on the flavor assumptions that are made in the fermion sector\cite{Almeida:2021asy,Bissmann:2020mfi,Ethier:2021bye,Ellis:2020unq}.  In this note, we generalize our previous results for EWPO to include flavor effects from the 4-fermion operators involving at least 2 quarks that contribute at one-loop level.  We  present some interesting numerical consequences of allowing the 4-fermion operators to have an
arbitrary flavor structure and compare our results with those from fits to top quark data at the LHC. 

\section{Basics  }
\label{sec:basics}

The SMEFT
parameterizes new physics through an expansion in higher dimensional operators\cite{Brivio:2017vri},
\begin{equation}
{\cal L}={\cal L}_{SM}+\Sigma_{k=5}^{\infty}\Sigma_{\alpha=1}^n {{C}_\alpha^k\over \Lambda^{k-4}} \Op_\alpha^k\, ,
\label{eq:lsmeft}
\end{equation}
where the $SU(3)\times SU(2)_L\times U(1)_Y$ invariant 
 operators, $O_\alpha^k$,  are constructed from SM fields and all of  the effects of the BSM
  physics  reside in the coefficient functions, $C_\alpha^k$. 
We use the Warsaw basis \cite{Buchmuller:1985jz,Grzadkowski:2010es}, assume all coefficients are real, include only dimension-6 operators
and do not consider CP violation.  Observables are computed consistently to 
${\cal{O}}({M_Z^2\over \Lambda^2}) $, corresponding to only one operator insertion at the amplitude level,  including all NLO QCD and electroweak contributions, as described in 
Ref. \cite{Dawson:2019clf}.   At lowest order (LO), only the operators  ${\Op_{ll}},
  {\Op}_{\phi W B}, 
 \Op_{\vp D},
    {\Op}_{\phi e},
    {\Op}_{\phi u},
      {\Op}_{\phi d},
             {\Op}_{\phi q}^{(3)},
               {\Op}_{\phi q}^{(1)},
   {\Op_{\vp l}^{(3)}},$ and 
   ${\Op}_{\phi l}^{(1)}$
   contribute. 
 Ref. \cite{Dawson:2019clf} also included the contributions of the operators
\begin{eqnarray}&&
{\Op}_{ed}\, ,{\Op}_{ee}\, ,{\Op}_{eu}\, ,{\Op}_{lu}\,,{\Op}_{ld}\, ,{\Op}_{le}\, ,{\Op}_{lq}^{(1)}\, ,{\Op}_{lq}^{(3)}\, ,{\Op}_{\phi B}\, ,{\Op}_{\phi W}
\, ,{\Op}_{\square},
\nonumber \\
&&{\Op}_{qe}\, ,{\Op}_{uB}\, ,{\Op}_{uW}\, ,{\Op}_{W}\,, {\Op}_{qd}^{(1)}\,, {\Op}_{qq}^{(3)}\,,{\Op}_{qq}^{(1)}\,,{\Op}_{qu}^{(1)}\,,{\Op}_{ud}^{(1)}\,
,{\Op}_{uu}\,\,, {\Op}_{dd} \, ,
\end{eqnarray}
that first arise at NLO, 
where the coefficients of the 2-quark and 4-quark operators were assumed to be flavor independent.  All 
renormalization group mixing of the operators was included, as required to obtain a finite NLO result. 
Contrary to Ref. \cite{Dawson:2019clf}, here we include the effects of arbitrary flavor interactions between quarks in the 4-fermion operators. 
In doing our computation, we set the SM CKM  matrix to be the unit matrix and we assumed that the 2-fermion operators that appear at lowest order are diagonal, but not proportional to unity.  For example\footnote{After renormalization, we drop the generation indices on the 2-fermion operators.Therefore, the results presented here differ from those of  Ref. \cite{Dawson:2019clf}  only for  the 4-quark  and 2-quark, 2-lepton operators listed in Table \ref{tab:opdef}. }
   \begin{equation}
   {\Op}_{\phi u,[11]}\ne {\Op}_{\phi u,[22]}\ne {\Op}_{\phi u,[33]}\, \quad \quad{\Op}_{\phi u,[12]}={\Op}_{\phi u,[13]}={\Op}_{\phi u,[23]}={\Op}_{\phi u,[13]}=0\, .
   \label{eq:assum}
      \end{equation} 
It is worth noting that setting the SM CKM matrix to 1 is enough to ensure that at LO only the diagonal terms of the 2-fermion operators contribute to all Z-pole observables. At NLO, the analysis becomes more complicated.  However, since we are interested only in the contributions induced by the 4-fermion operators, it is straightforward to observe that our choice of taking the SM CKM matrix to be unity  affects our results mostly through the renormalization group (RGE)  mixing of operators. The only observable where this is not true is the W width where extra terms proportional to the off-diagonal terms of the SM CKM matrix are not included. For these reasons, at the level we are working, we expect the effects 
on our results for  the $4-$fermion interactions from our choice of taking the SM CKM matrix to be unity to be small.

Defining the fermion fields as,
\begin{equation}
q_{L,r}=\left(
\begin{array}{l}u_{L,r}\\ d_{L,r}\end{array}\right), ~ u_{R,r}, ~ d_{R,r}, ~
l_{L,r}=\left(\begin{array}{l}\nu_{L,r}\\ e_{L,r}\end{array}\right),~ e_{R,r} ~,
\end{equation}
where $r=1,2,3$ is a generation index, we considered the operators in  Table \ref{tab:opdef}, where we drop the lepton flavor indices. 
\begin{table}[t]
\centering
\renewcommand{\arraystretch}{1.5}
\begin{tabular}{||c|c||c|c||}
\hline\hline
\multicolumn{4}{||c||}{2-quark 2-lepton operators}\\
\hline
${\Op}_{lq,[tp]}^{(1)}$ 
& $({\overline l}_{L}\gamma_\mu l_{L})({\overline q}_{L,t}\gamma^\mu q_{L,p})$
& ${\Op}_{lq,[tp]}^{(3)}$ 
& $({\overline l}_{L}\gamma_\mu \sigma^A l_{L})({\overline q}_{L,t}\gamma^\mu \sigma^Aq_{L,p})$\\
\hline
${\Op}_{eu,[tp]}$ 
& $({\overline e}_{R}\gamma_\mu e_{R})({\overline u}_{R,t}\gamma^\mu u_{R,p})$ 
& ${\Op}_{ed,[tp]}$ 
& $({\overline e}_{R}\gamma_\mu e_{R})({\overline d}_{R,t}\gamma^\mu d_{R,p})$ \\
\hline
 ${\Op}_{lu,[tp]}$ 
 & $({\overline l}_{L}\gamma_\mu l_{L})({\overline{u}}_{R,t}\gamma^\mu u_{R,p})$
& ${\Op}_{ld,[tp]}$ 
&$({\overline l}_{L}\gamma_\mu l_{L})({\overline d}_{R,t}\gamma^\mu d_{R,p})$ \\
\hline
$  {\Op}_{qe,[rs]}$
&$({\overline q}_{L,r}\gamma_\mu q_{L,s})({\overline e}_{R}\gamma^\mu e_{R})$&&\\
\hline
\multicolumn{4}{||c||}{4-quark operators}\\
\hline
${\Op}_{qq,[rstp]}^{(1)}$ 
& $({\overline q}_{L,r}\gamma_\mu q_{L,s})({\overline q}_{L,t}\gamma^\mu q_{L,p})$
& $ {\Op}_{qq,[rstp]}^{(3)}$ 
&$({\overline q}_{L,r}\gamma_\mu \sigma^A q_{L,s})({\overline q}_{L,t}\gamma^\mu \sigma^A q_{L,p})$ \\
\hline
${\Op}_{uu,[rstp]}$ 
& $({\overline u}_{R,r}\gamma_\mu u_{R,s})({\overline u}_{R,t}\gamma^\mu u_{R,p})$
&${\Op}_{dd,[rstp]}$
& $({\overline d}_{R,r}\gamma_\mu d_{R,s})({\overline d}_{R,t}\gamma^\mu d_{R,p})$\\
\hline
${\Op}_{ud,[rstp]}^{(1)}$
&$({\overline u}_{R,r}\gamma_\mu u_{R,s})({\overline d}_{R,t}\gamma^\mu d_{R,p})$
&${\Op}_{qu,[rstp]}^{(1)}$ 
&$({\overline q}_{L,r}\gamma_\mu q_{L,s})({\overline u}_{R,t}\gamma^\mu u_{R,p})$\\
\hline
${\Op}_{qd,[rstp]}^{(1)}$
& $({\overline q}_{L,r}\gamma_\mu q_{L,s})({\overline d}_{R,t}\gamma^\mu d_{R,p})$&&\\
 \hline \hline
\end{tabular}
\caption{Dimension-6  $4$ -fermion operators  contributing to  $Z$ and $W$ pole observables  at 
NLO QCD and electroweak order \cite{Dawson:2019clf} where $[r,s,t,p]=1,2,3$ are quark generation indices and
$\sigma^A$ are the Pauli matrices. 
\label{tab:opdef}}
\end{table}

\begin{figure}
  \centering
\includegraphics[width=.4\textwidth]{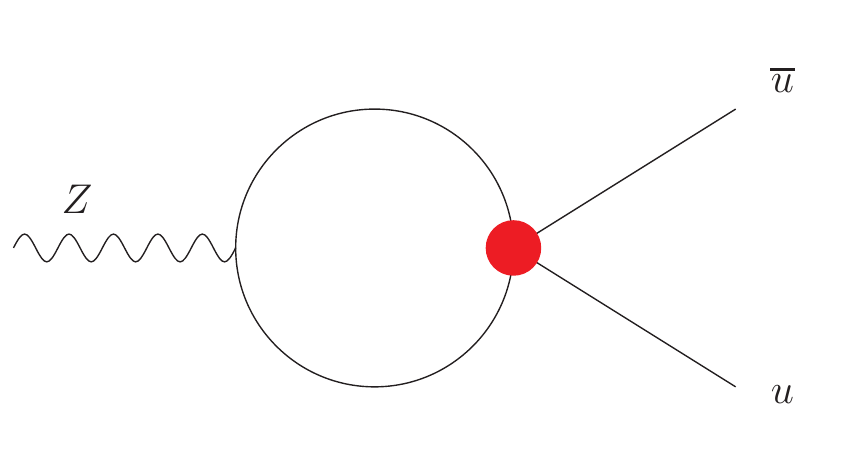}
 \caption{Sample   diagram containing $4-$ fermion operators contributing to $Z\rightarrow u {\overline {u}}$ at NLO in the SMEFT.   The fermions in the loop can be any quark or lepton (heavy or light).  The red  circle represents insertions of the operators of Table \ref{tab:opdef}.  \label{fig:diag}}
\end{figure}

\section{Results}
\label{sec:zpole}

The observables we consider are:
\begin{eqnarray}
&&M_W, \Gamma_W, \Gamma_Z,  \sigma_h,  R_l, A_{l,FB}, R_b, R_c,  A_{FB,b}, A_{FB,c},  A_b, A_c, A_l\, .
\label{eq:quan}
\end{eqnarray} 
The SM results for these observables are quite precisely known and we use the experimental and theoretical SM results shown in Table III of Ref. \cite{Dawson:2019clf}. 
The NLO SMEFT results for the observables of Eq. \ref{eq:quan} contain one-loop
contributions from the dimension-6 operators  of Table \ref{tab:opdef}
and the  full electroweak and QCD NLO amplitudes assuming that the flavor interactions are independent of fermion generation are
in the supplemental material of Ref. \cite{Dawson:2019clf}.    
Here we focus on the effects of the $4$-fermion operators for on-shell 2-body $Z$ and $W$ decays such as those
 shown in Fig. \ref{fig:diag} and  allow for an arbitrary flavor dependence in the $4-$fermion operators.  
 When the internal fermions are top quarks, contributions that are enhanced by factors of $M_t^2/M_Z^2$ arise. 
 Such contributions contribute to $Z\rightarrow b {\overline{b}}$  generically through the coefficients, $C_{\alpha,[3333]}$, and  to $Z\rightarrow f_i {\overline{f}}_i $ (where $f_i$ is a light fermion) through the coefficients, $C_{\alpha, [33ii]}$, etc.
  We note
 that not all combinations of generation indices arise in the NLO calculation of the EWPO.  For example,
 the operator $\Op_{qq}^{(1)}$ occurs with $i, i^\prime =1,2$, (where $i\ne i^\prime$) 
 \begin{equation}
 C_{qq,[3333]}^{(1)},~C_{qq,[33ii]}^{(1)}=C_{qq,[ii33]}^{(1)},~C_{qq,[3ii3]}^{(1)}=C_{qq,[i33i]}^{(1)},~
 C_{qq,[iii^\prime i^\prime ]}^{(1)},~C_{qq,[ii^\prime i^\prime i]}^{(1)}\,~
 C_{qq,[iii i]}^{(1)}\, .
 \end{equation}
 In our calculation we never encounter operators with more than 2 different flavor indices, due to our choice of flavor structure.

The  SMEFT predictions for the observables are,
\begin{eqnarray}
O_i^{SMEFT, LO}&=&O_i^{SM,LO}+\delta O_i^{LO}(C_j)\nonumber  \\
O_i^{SMEFT,NLO}&=&O_i^{SM,NLO}+\delta O_i^{NLO}(C_j)\, .
\label{eq:defs}
\end{eqnarray}
We present numerical results for the observables of Eq.
\ref{eq:quan} in the supplemental material attached to this note.  
In the limit where  the $C_{\alpha,[rstp]}$ are independent of the generation indices, our previous result is recovered. 

We perform a $\chi^2$ fit to the EWPO data.  As an example,  
if only the $3^{rd}$ generation quarks contribute to the $4$- fermion operators, the NLO contribution 
to the $\chi^2$  is\footnote{$\Delta\chi^2_{[3333]} $ must be added to the full SMEFT result including all other operators. }

\begin{eqnarray}
\Delta \chi^2_{[3333]}&=& 
-0.45893 {C_{ud,[3333]}^{(1)}}
  +1.0712 C_{qu,[3333]}^{(1)}
    -0.16008 {C_{qq,[3333]}^{(3)}}
  -1.7914 {C_{qq,[3333]}^{(1)}}
   \nonumber \\ &&
 +0.37024 {C_{qd,[3333]}^{(1)}}
  +0.019645 {C_{dd,[3333]}} +{\vec{X}}^TM_{[3333]}{\vec{X}}\, ,
  \label{eq:ch330}
   \end{eqnarray}
   with ${\vec{X}}^T=({C_{ud,[3333]}^{(1)}}, C_{qu,[3333]}^{(1)},
    {C_{qq,[3333]}^{(3)}}, {C_{qq,[3333]}^{(1)}},
 {C_{qd,[3333]}^{(1)}}, {C_{dd,[3333]}})$ and 
 \begin{equation}
M_{[3333]}=\left( \begin{array}{cccccc}
 0.039703 & 
 -0.38683 &
 +0.057807&
 +0.6469&
 -0.060256&
 -0.0033991 \\
  & 
 1.1917 &
 -0.35618 &
 -3.9858 &
 +0.28411 &
 +0.016559 \\
 &&
 0.026614 &
 +0.59564&
 -0.042458 &
  -0.0024745
  \\
 &&&
     3.3328 &
 -0.47512 &
  -0.027691
  \\
 &&&&
    0.022951 &
 0.0025794 
 \\
 &&&&&
   7.2752 \times 10^{-5}
   \end{array}
  \right) \, .
  \label{eq:ch33}
 \end{equation}
 It is apparent that the largest sensitivity in this case is to $C_{qq,[3333]}^{(1)}$ and to 
 $C_{qu,[3333]}^{(1)}$.
  \begin{table}[t]
\centering
\renewcommand{\arraystretch}{1.5}
\begin{tabular}{||c|c|c|c|c||}
\hline\hline
& No flavor dep. & Only $3^{rd}$ gen.& At least $3^{rd}$ gen. & Only $1^{st}, 2^{nd}$ gen. \\
\hline \hline
${C_{qq}^{(1)}\over \Lambda^2}$ &[-0.93,1.48] & [-0.80,1.34]& [-0.93,1.54]& [-11.45,10.81]\\
\hline
${C_{qq}^{(3)}\over \Lambda^2}$ &[-0.32,0.29] &[-9.01,15.02] & [-0.49,0.42]& [-0.94,.0.88] \\
\hline
${C_{qu}^{(1)} \over \Lambda^2}$ &[-2.17,1.33] & [-2.24,1.34] & [-2.20,1.35]& [-46.61,42.53] \\ \hline
${C_{qd}^{(1)} \over \Lambda^2}$ &[-9.76,4.98] &[-21.00.,4.87] & [-8.98,4.79]& [-85.71,90.22]\\ \hline
${C_{uu} \over \Lambda^2}$ &[-1.14,0.99] & -& [-1.20,1.04] & [-22.56,19.48] \\ \hline
${C_{dd}\over \Lambda^2} $ &[-50.60,26.29] &[-364.80,94.77] & [-89.93,31.77]& [-87.75,82.43]\\ \hline
${C_{ud}^{(1)} \over \Lambda^2}$ &[-3.01,5.62] & [-4.06,15.62] & [-3.17,6.21]& [-45.74,50.83] \\ \hline
${C_{lq}^{(1)}\over \Lambda^2}$ &[-0.25,0.66] & [-0.24,0.63] & [-0.24,0.63]&[-13.25,5.44] \\ \hline
${C_{lq}^{(3)} \over \Lambda^2}$ &[-0.32,0.57] & [-0.29,0.68] & [-0.29,0.68]& [-1.92,1.02] \\ \hline
${C_{lu} \over \Lambda^2}$ &[-0.49,0.19] & [-0.53,0.21] & [-0.53,0.21]& [-6.62,2.74] \\ \hline
${C_{ld} \over \Lambda^2}$ &[-3.76,8.71] & [-11.99,25.13] & [-11.99,25.13]& [-5.45,13.28] \\ \hline
${C_{qe}\over \Lambda^2} $ &[-0.75,0.48] & [-0.72,0.45] & [-0.72,0.45]& [-10.50,17.30] \\ \hline
${C_{ed}\over \Lambda^2} $ &[-11.80,6.71] & [-36.23,18.28] & [-36.23,18.28]& [-17.41,10.52]\\ \hline
${C_{eu}\over \Lambda^2} $ &[-0.36,0.58] & [-0.39,0.62] & [-0.39,0.62]& [-5.23,8.72] \\ 
\hline
\hline
\end{tabular}
\caption{
$95\%$ CL single parameter limits  in $TeV^{-2}$ from EWPO in the Warsaw basis.   The second column corresponds to the NLO results in our previous paper with no flavor dependence in the quark and lepton sectors, the $3^{rd}$ column has the only non-zero contributions from the $3^{rd}$ generation fermions,
 the $4^{th}$ column sets coefficients which only involve the $1^{st}$ and $2^{nd}$ generation quarks to 0, and the $5^{th}$ column 
has equal contributions for each coefficient from operators involving the $1^{st}$ and $2^{nd}$ generations, with no contributions from the $3^{rd}$ generation. }
\label{tab:ex}
\end{table}
 
\begin{figure}
  \centering
\includegraphics[width=.6\textwidth]{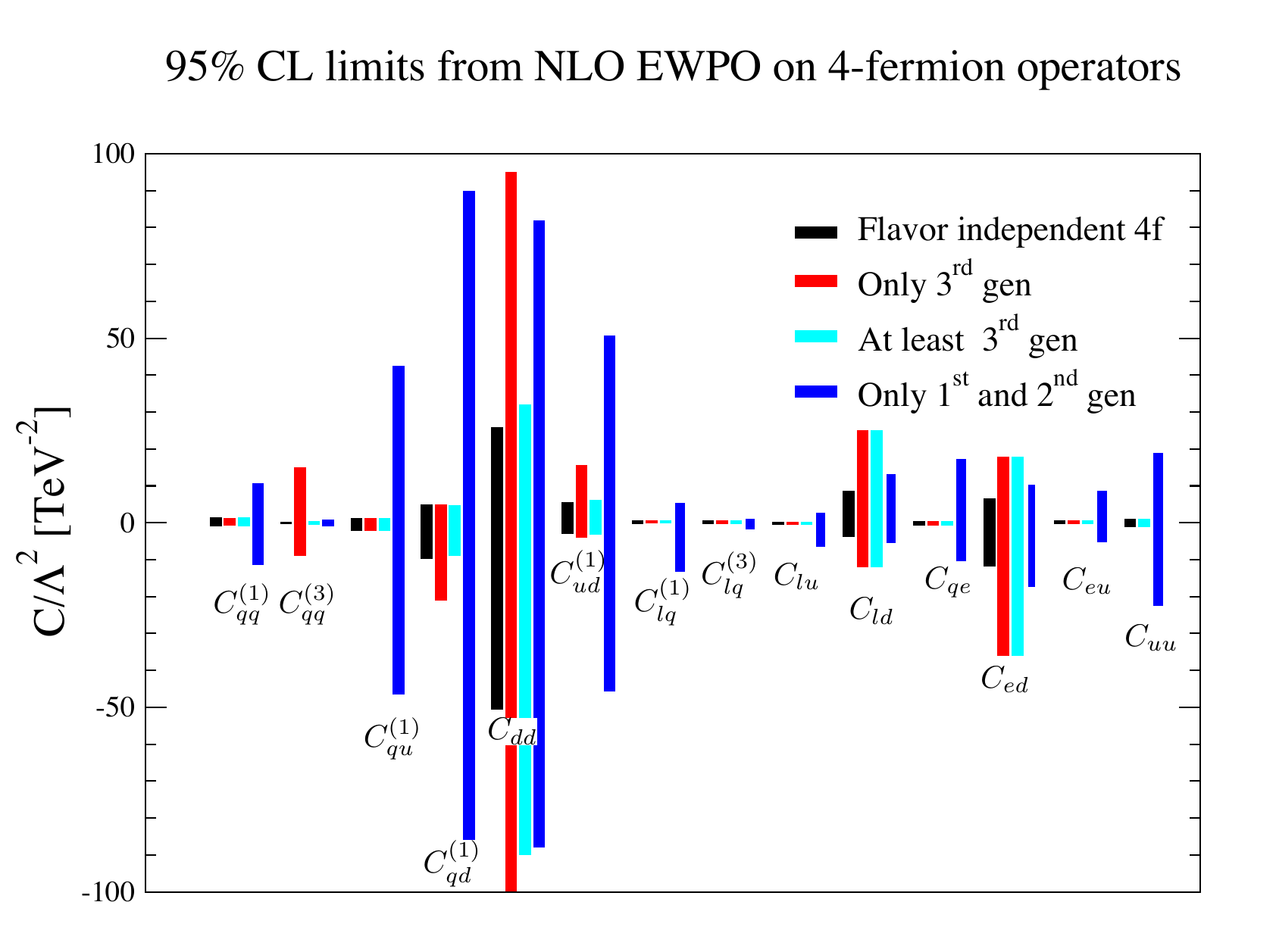}
 \caption{ Comparison of single parameter limits from loop corrections to EWPO with different flavor assumptions described in the text. 
  {\label{fig:3bar}}}
\end{figure}

 We show the $95\%$ confidence level single parameter limits on the $4-$fermion operators of Table \ref{tab:opdef} with various flavor assumptions in Table \ref{tab:ex}. 
 All of our coefficients are evaluated at a scale $M_Z$. Column 2 is the result of Ref. \cite{Dawson:2019clf} where there is no flavor dependence in the $4-$fermion coefficients,
 \begin{eqnarray}
 \text{Column 2 of Tab. \ref{tab:ex}:} \quad 
 C_{\alpha,[3333]}&=&C_{\alpha,[33ii]}=C_{\alpha,[ii33]}=C_{\alpha,[3ii3]}=C_{\alpha,[i33i]}\nonumber \\ &=&C_{\alpha,[iiii ]}=
 C_{\alpha,[iii^\prime i^\prime ]}=C_{\alpha,[ii^\prime i^\prime i]}\, .
 \end{eqnarray} 
 It is of interest to consider the scenario where the $3^{rd}$ generation $4-$fermion operators are different from those of generations $i=1,2$.   Column 3 assumes
 that the $4-$fermion operators are only generated for the $3^{rd}$ generation quarks and can be obtained from
 Eqs. \ref{eq:ch330} and \ref{eq:ch33},
 \begin{eqnarray}
 \text{Column 3 of Tab. \ref{tab:ex}:} 
 && C_{\alpha,[3333]}\ne 0,\quad C_{\alpha,[33ii]}=C_{\alpha,[ii33]}=C_{\alpha,[i33i]}= C_{\alpha,[3ii3]}=0
 \nonumber \\ && 
 C_{\alpha,[iii^\prime i^\prime ]}=C_{\alpha,[ii^\prime i^\prime i]}=\C_{\alpha,[iiii]}=0\, .
 \end{eqnarray}
 The $4^{th}$ column assumes that $4-$fermions operators where only $1^{st}$ and $2^{nd}$ generation quarks appear are zero 
 and the remaining operators are equal,
 \begin{eqnarray}
 \text{Column 4 of Tab. \ref{tab:ex}:} 
 && C_{\alpha,[3333]}=C_{\alpha,[33ii]}=C_{\alpha,[ii33]}=
 C_{\alpha,[3ii3]}=C_{\alpha,[i33i]}\ne 0
 \nonumber \\  && 
 C_{\alpha,[iii^\prime i^\prime ]}=C_{\alpha,[ii^\prime i^\prime i]}
 =C_{\alpha,[iiii]}=0\, .
 \end{eqnarray}
 Finally, the $5^{th}$ column assumes that there are no contributions to 
 $4-$fermion operators from $3^{rd}$ generation quarks and that all the coefficients involving $1^{st}$
 and $2^{nd}$ generation quarks for each operator are equal,
 \begin{eqnarray}
 \text{Column 5 of Tab. \ref{tab:ex}:} &&
  C_{\alpha,[3333]}=C_{\alpha,[33ii]}=
 C_{\alpha,[ii33]}=C_{\alpha,[3ii3]}=C_{\alpha,[i33i]}= 0\, ,
 \nonumber \\ &&
 C_{\alpha,[ii i^\prime i^\prime]}=
 C_{\alpha,[i i^\prime i^\prime i]}
 =C_{\alpha,[iiii]}\ne  0\, .
 \end{eqnarray}
 It is obvious that the results are extremely sensitive to the flavor assumptions and that there are large cancellations between contributions with massive
 internal top quarks and the massless fermions.  We plot these results in Fig. \ref{fig:3bar}.
\begin{figure}
  \centering
\includegraphics[width=.6\textwidth]{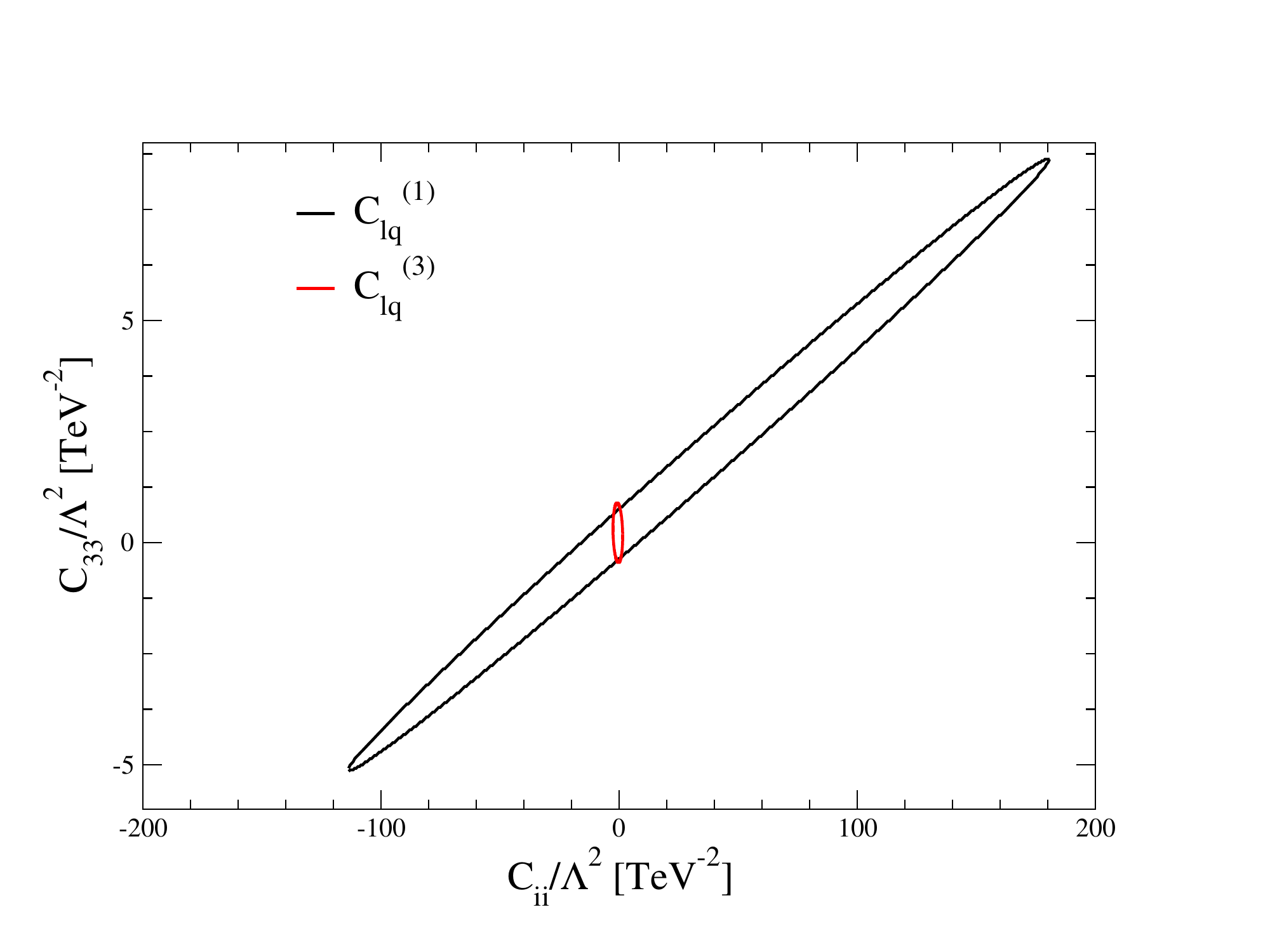}
 \caption{ Fit to EWPO with the 
 only non-zero Wilson coefficients  being $C_{lq}^{(1)}$ and $C_{lq}^{(3)}$.  The quark generation index $i=1,2$.}
 \label{fig:diag2}
\end{figure}
\begin{figure}
  \centering
\includegraphics[width=.4\textwidth]{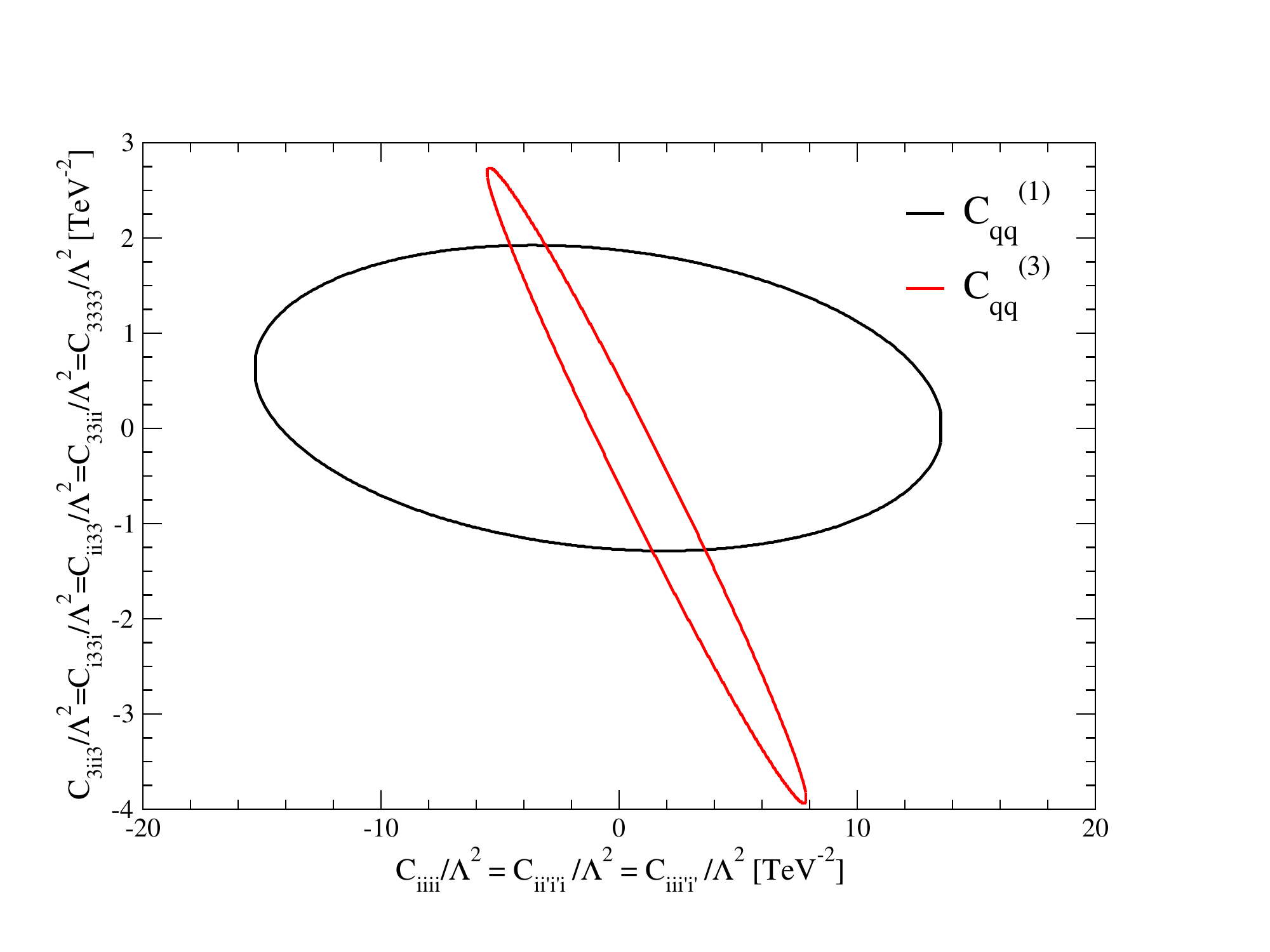}
\includegraphics[width=.4\textwidth]{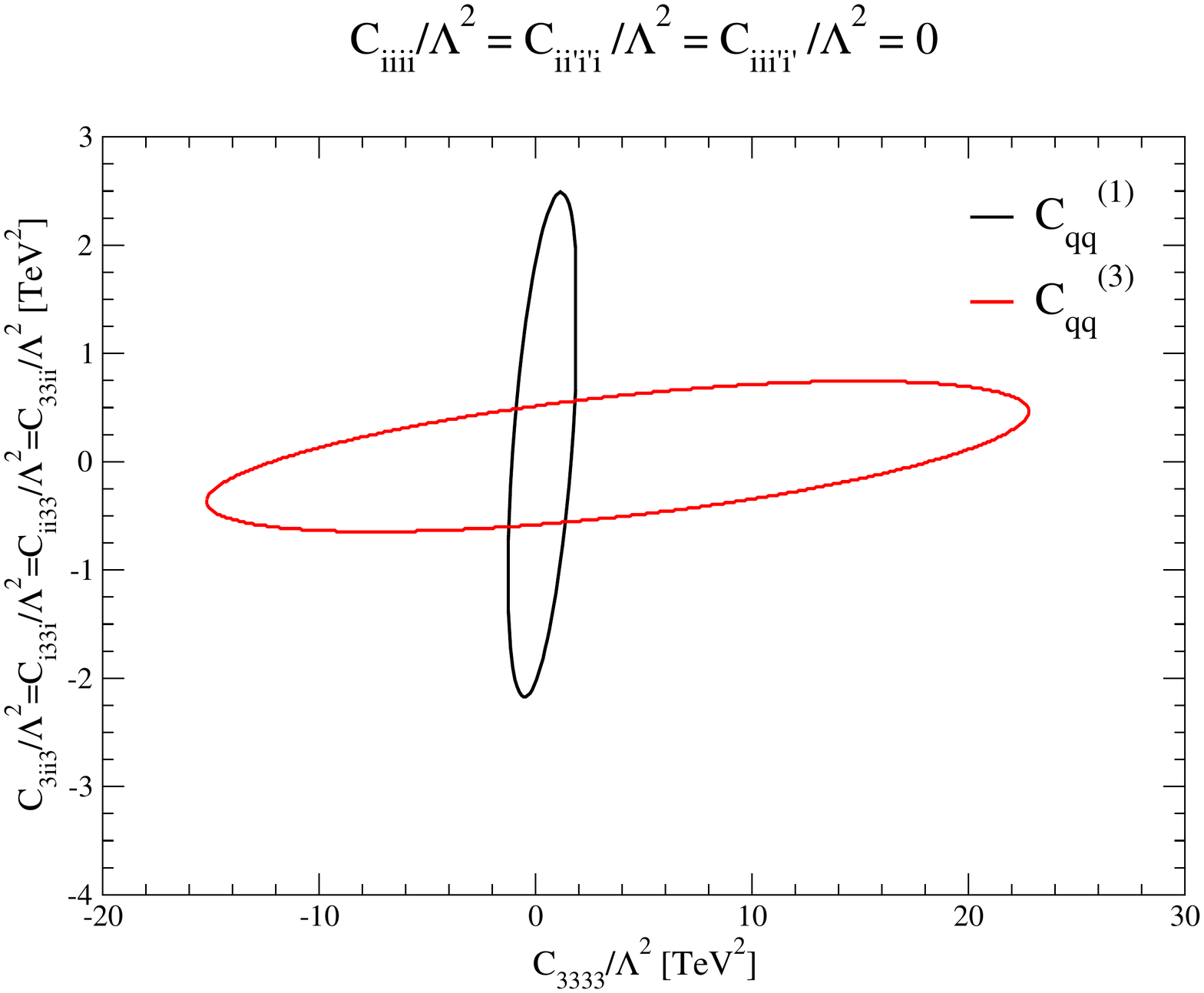}
 \caption{Fit to EWPO with  the only non-zero Wilson coefficients  $C_{qq}^{(1)}$ and $C_{qq}^{(3)}$.  The quark generation index $i=1,~2$ and $i \ne i^\prime$. {\label{fig:diag3}}}
\end{figure}
  Figs. \ref{fig:diag2} and \ref{fig:diag3}  show several $2$- parameter fits with various assumptions about the flavor structure of the $4$-fermion operators.  Again, the numerical values of the fits are depend on the generation structure assumed for the 4-fermion operators and strong correlations are observed.

\begin{figure}
  \centering
\includegraphics[width=.6\textwidth]{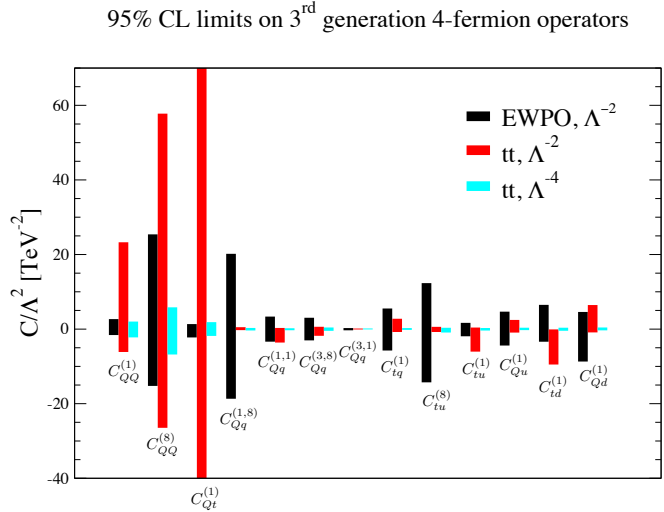}
 \caption{ Comparison of single parameter limits from loop corrections to EWPO
  involving $3^{rd}$ generation $4-$ fermion interactions with similar limits from LHC $t {\overline{t}}$ 
  production\cite{Zhang:2017mls}.
  {\label{fig:3g}}}
\end{figure}

  In Table \ref{tab:two}, we compare the limits from the single parameter EWPO NLO fits to those obtained from fits to LHC $t {\overline{t}}$ data\cite{ethier2021combined,Zhang:2017mls,Brivio:2019ius,Buckley:2015nca}, using the notation of Ref. \cite{Aguilar-Saavedra:2018ksv}.  Our NLO fits are only consistent to ${\cal{O}}({1\over\Lambda^2})$, and extending them to ${\cal{O}}({1\over\Lambda^4})$ would require double insertions of dimension-6 operators and  the inclusion of dimension-8 operators.   Comparing the ${\cal{O}}({1\over\Lambda^2})$ EWPO and  the $t{\overline{t}}$ results, we see that for several operators the EWPO result is comparable or better than the top quark result.  We note, however, that most of the power of the $t {\overline{t}}$ fit  is coming at 
  ${\cal{O}}({1\over\Lambda^4})$. It is amusing to note that the EWPO results for $C_{QQ}^{(1)}$ and  $C_{Qt}^{(1)}$ are still relevant even when compared to the $t {\overline{t}}$ ${\cal{O}}({1\over\Lambda^4})$ results.
The impact of the EWPO NLO data is illustrated graphically in Fig. \ref{fig:3g} and suggests that including the NLO EWPO result with non-universal flavor effects in the global fits could have
  a significant effect. 
  
\begin{figure}
  \centering
\includegraphics[width=.6\textwidth]{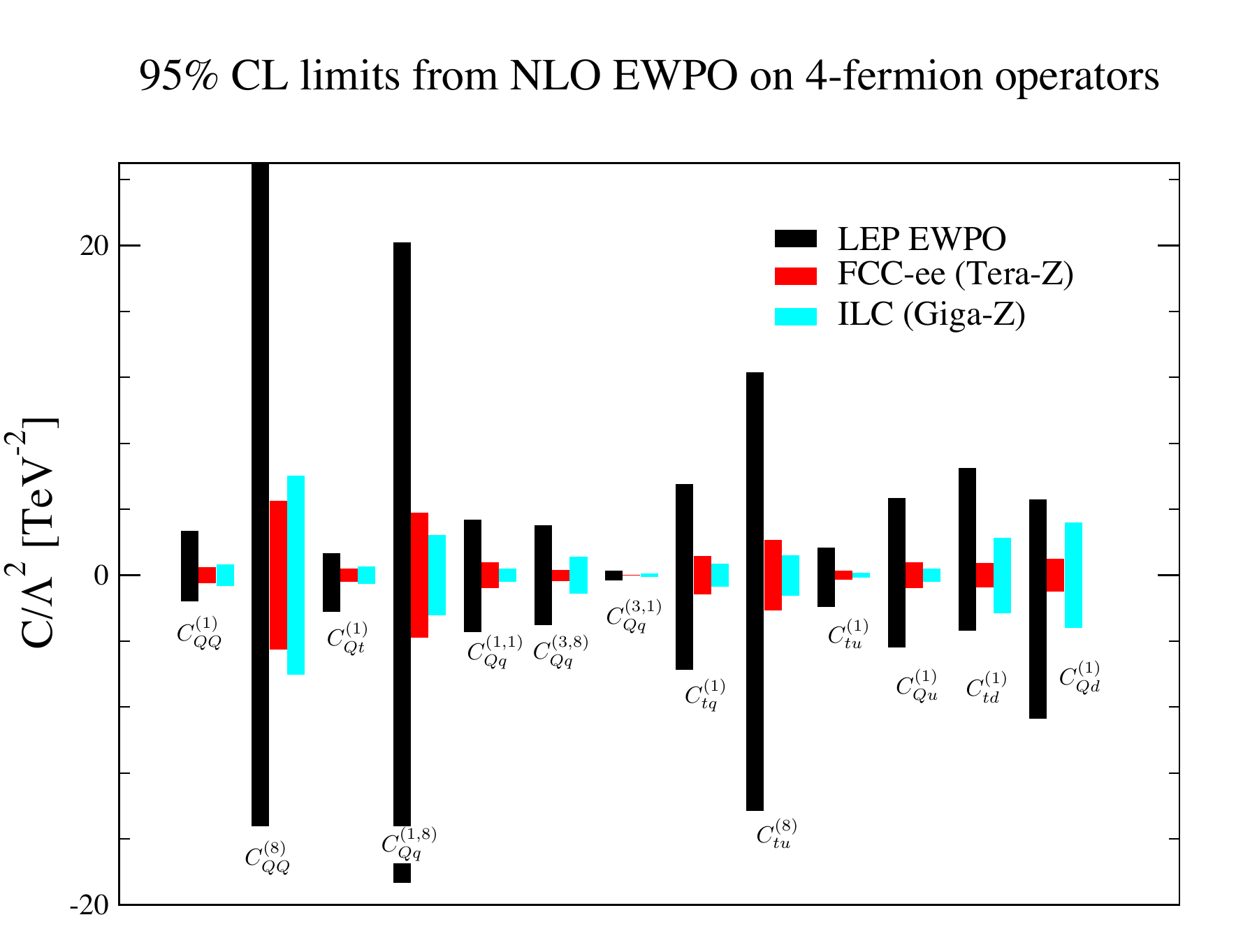}
 \caption{ Comparison of single parameter limits from loop corrections to EWPO
  involving $3^{rd}$ generation $4-$ fermion interactions with projected Z pole limits from a Tera-Z program at the
  FCC-ee\cite{Blondel:2018mad} and with a Giga-Z ILC run\cite{fujii2019tests} .
  {\label{fig:fcc}}}
\end{figure}

 In Fig. \ref{fig:fcc}, we compare the current precision from the EWPO on the $3^{rd}$ generation operators with that projected from a Tera-Z  run at FCC-ee with an assumed integrated luminosity 
 of $150~ab^{-1}$  ($3\times 10^{12}$ visible Z's) and with a Giga-Z run at the ILC with an integrated luminosity of $100~fb^{-1}$ ($10^{9}$ Z's).  This figure assumes that current theory uncertainties are halved, assumes the FCC-ee precision of Table II in Ref. \cite{Blondel:2018mad}
 and the ILC Giga-Z numbers of Table 9  in Ref. \cite{fujii2019tests}.  Due to the polarization, for some observables the projected ILC precision surpasses that of the FCC-ee, despite the smaller assumed luminosity.

\begin{table}[t]
\centering
\renewcommand{\arraystretch}{1.5}
\begin{tabular}{||c|c|c|c||}
\hline\hline
Operator & EWPO & $t {\overline t}(1/\Lambda^2)$ & $t {\overline t}(1/\Lambda^4)$ \\
\hline \hline
${C_{QQ}^{(1)}\over \Lambda^2}\equiv 2{C_{qq,[3333]}^{(1)}\over \Lambda^2}-\frac23 {C_{qq,[3333]}^{(3)}\over \Lambda^2} $ &[-1.61,2.68]  & [-6.132, 23.281] & [-2.229,2.019]\\ \hline
$ {C_{QQ}^{(8)}\over \Lambda^2}\equiv 8 {C_{qq,[3333]}^{(3)}\over \Lambda^2} $ &[-15.23,25.41] &[-26.471,57.778] &[-6.812,5.834]\\ \hline
${C_{Qt}^{(1)}\over \Lambda^2}\equiv {C_{qu,[3333]}^{(1)}\over \Lambda^2}$  &[-2.24,1.35] &[-195,159] &[-1.830,1.862] \\ \hline
${C_{Qq}^{(1,8)}\over \Lambda^2}\equiv {C_{qq,[i33i]}^{(1)}\over \Lambda^2}+3 {C_{qq,[i33i]}^{(3)}\over \Lambda^2}$ &[-18.67,20.19] &[-0.273,0.509] &[-0.373,0.309] \\ \hline
${C_{Qq}^{(1,1)}\over \Lambda^2}\equiv {C_{qq,[ii33]}^{(1)}\over \Lambda^2}+{1\over 6}{C_{qq,[i33i]}^{(1)}\over \Lambda^2}+{1\over 2} {C_{qq,[i33i]}^{(3)}\over \Lambda^2}$  &[-3.47,3.36] &[-3.603,0.307] &[-0.303,0.225]\\ \hline
${C_{Qq}^{(3,8)}\over \Lambda^2}\equiv {C_{qq,[i33i]}^{(1)}\over \Lambda^2}-{C_{qq,[i33i]}^{(3)}\over \Lambda^2}$  &[-3.03, 3.04] &[-1.813,0.625] &[-0.470,0.439] \\ \hline
${C_{Qq}^{(3,1)}\over \Lambda^2}\equiv {C_{qq,[ii33]}^{(3)}\over \Lambda^2}+{1\over 6}\biggl({C_{qq,[i33i]}^{(1)}\over \Lambda^2}-{C_{qq,[i33i]}^{(3)}\over \Lambda^2}\biggr)$ &[-0.32, 0.28] &[-0.099,0.155] &[-0.088,0.166]\\ \hline
${C_{tq}^{(1)}\over \Lambda^2}\equiv {C_{qu,[ii33]}^{(1)}\over \Lambda^2}$  &[-5.74,5.52] &[-0.784,2.771] &[-0.205,0.271]\\ \hline
${C_{tu}^{(8)}\over \Lambda^2}\equiv 2 {C_{uu,[i33i]}\over \Lambda^2}$  &[-14.28,12.33] &[-0.774,0.607] &[-0.911,0.347]\\ \hline
${C_{tu}^{(1)}\over \Lambda^2}\equiv {C_{uu,[ii33]}\over \Lambda^2}+\frac13 {C_{uu,[i33i]}\over \Lambda^2}$  &[-1.93,1.67] &[-6.046,0.424] &[-0.380,0.293]\\ \hline
${C_{Qu}^{(1)}\over \Lambda^2}\equiv {C_{qu,[33ii]}^{(1)}\over \Lambda^2}$  &[-4.40,4.67] &[-0.938,2.462] &[-0.281,0.371]\\ \hline
${C_{td}^{(1)}\over \Lambda^2}\equiv {C_{ud,[33jj]}^{(1)}\over \Lambda^2}$   &[-3.38,6.50] &[-9.504,-0.086] &[-0.449,0.371]\\ \hline
${C_{Qd}^{(1)}\over \Lambda^2}\equiv {C_{qd,[33jj]}^{(1)}\over \Lambda^2}$  &[-8.70,4.59] &[-0.889,6.459] &[-0.332,0.436]\\ \hline
\hline \hline
\end{tabular}
\caption{$95\%$ CL single parameter limits in $TeV^{-2}$ from NLO contributions of $3^{rd}$ generation 4-fermion operators to EWPO ($1^{st}$ column) to LHC $t {\overline {t}}$ results $t {\overline {t}}$ results using the ${\cal{O}}({1\over \Lambda^2}) $ and ${\cal{O}}({1\over \Lambda^4})$ expansions\cite{ethier2021combined}. The generation index $i=1,2$ and $j=1,2,3$ and the scale $\Lambda=1~TeV$.}
\label{tab:two}
\end{table}

\section{Conclusions}
\label{sec:conc}
We have included flavor non-universal effects from $4-$fermion operators with at least 2 quarks into the NLO electroweak and QCD corrections to the SMEFT predictions for the precision electroweak
observables.  Our results are presented in a numerical form that can  be incorporated in the global fitting programs and suggest that the flavor assumptions on the $4-$ fermion operators  can have a significant effect. 
In particular we showed that the bounds obtained from EWPO on the $C_{QQ}^{(1)}$ and $C_{Qt}^{(1)}$ operators, that appear in the EWPO only at NLO, are competitive with respect to those obtained from current LHC $t {\overline {t}}$ observables. 
Numerical results are posted at
\url{https://quark.phy.bnl.gov/Digital_Data_Archive/dawson/ewpo_22}.
 
 \section*{Acknowledgements}
We thank  Susanne Westhoff, Danny van Dyk, and Sebastian Bruggisse for pointing out that including non-trivial flavor dependencies   in the SMEFT
NLO predictions for the EWPO
could be important for the global fits and also for valuable comments on the manuscript. 
S.D.   is  supported by the U.S. Department of Energy under Grant Contract  de-sc0012704.  
The work of PPG has received financial support from Xunta de Galicia (Centro singular de investigaci\'on de Galicia accreditation 2019-2022), by European Union ERDF, and by  ``Mar\'ia  de Maeztu"  Units  of  Excellence program  MDM-2016-0692  and  the Spanish Research State Agency.

\bibliographystyle{utphys}
\bibliography{mw}

\end{document}